\documentclass[conference]{IEEEtran}
\IEEEoverridecommandlockouts
\usepackage{cite}
\usepackage{amsmath,amssymb,amsfonts}
\usepackage{algorithmic}
\usepackage{graphicx}
\usepackage{textcomp}
\usepackage{xcolor}
\usepackage{mathtools}
\usepackage{subcaption}
\usepackage{multicol,multirow}
\usepackage{enumerate,textcomp,array}
\usepackage{placeins}
\usepackage{threeparttable}
\usepackage{dblfloatfix}
\usepackage{comment}
\usepackage{gensymb}
\usepackage{nicematrix}
\NiceMatrixOptions{cell-space-top-limit = 1pt,cell-space-bottom-limit = 1pt}
\usepackage{filecontents}
\usepackage{tikz, tikzscale, pgfplots}
\usetikzlibrary{calc}
\usepackage{url}

\usepackage{breakurl}
\usepackage{booktabs}

\newcommand{\R}{\mathbb{R}}
\newcommand{\rmd}{\mathrm{d}}

\newcommand{\bA}{\mathbf{A}}
\newcommand{\bB}{\mathbf{B}}

\newcommand{\bgam}{\mathbf{\gamma}}

\newcommand{\bone}{\mathbf{1}}

\newcommand{\bT}{\mathbf{T}}
\newcommand{\bzero}{\mathbf{0}}

\newcommand{\tbB}{\tilde{\mathbf{B}}}

\newcommand{\diag}{\mathrm{diag}}

\def\BibTeX{{\rm B\kern-.05em{\sc i\kern-.025em b}\kern-.08em
    T\kern-.1667em\lower.7ex\hbox{E}\kern-.125emX}}
\begin{document}

\title{Analytical Models of Frequency and Voltage in Large-Scale All-Inverter Power Systems\

}
\thispagestyle{empty}
\pagestyle{plain}

\author{\IEEEauthorblockN{Marena Trujillo\IEEEauthorrefmark{1}\IEEEauthorrefmark{2}\IEEEauthorrefmark{3}, 
Amir Sajadi\IEEEauthorrefmark{2}\IEEEauthorrefmark{3}, 
Jonathan Shaw\IEEEauthorrefmark{4}, and 
Bri-Mathias Hodge\IEEEauthorrefmark{1}\IEEEauthorrefmark{2}\IEEEauthorrefmark{3}\IEEEauthorrefmark{4}}
\IEEEauthorblockA{\IEEEauthorrefmark{1}Department of Electrical, Computer, and Energy Engineering, University of Colorado of Boulder,
Boulder, Colorado 80309}
\IEEEauthorblockA{\IEEEauthorrefmark{2}Renewable and Sustainable Energy Institute, University of Colorado of Boulder,
Boulder, Colorado 80309}
\IEEEauthorblockA{\IEEEauthorrefmark{3}National Renewable Energy Laboratory, 
Golden, Colorado 80401}
\IEEEauthorblockA{\IEEEauthorrefmark{4}Department of Applied Mathematics, University of Colorado of Boulder,
Boulder, Colorado 80309\\
\{marena.trujillo, amir.sajadi, brimathias.hodge, jonathan.shaw\}@colorado.edu}
}

\maketitle

\begin{abstract}
Low-order frequency response models for power systems have a decades-long history in optimization and control problems such as unit commitment, economic dispatch, and wide-area control. With a few exceptions, these models are built upon the Newtonian mechanics of synchronous generators, assuming that the frequency dynamics across a system are approximately homogeneous, and assume the dynamics of nodal voltages for most operating conditions are negligible, and thus are not directly computed at all buses. As a result, the use of system frequency models results in the systematic underestimation of frequency minimum nadir and maximum RoCoF, and provides no insight into the reactive power-voltage dynamics. This paper proposes a low-order model of both frequency and voltage response in grid-forming inverter-dominated power systems. The proposed model accounts for spatial-temporal variations in frequency and voltage behavior across a system and as a result, demonstrates the heterogeneity of frequency response in future renewable power systems. Electromagnetic transient (EMT) simulations are used to validate the utility, accuracy, and computational efficiency of these models, setting the basis for them to serve as fast, scalable alternatives to EMT simulation, especially when dealing with very large-scale systems, for both planning and operational studies.

\end{abstract}


\label{sec:intro}

As the climate crisis drives extreme temperature changes, intensifying storms, and habitat destruction, the need for decarbonization is more urgent than ever. As a result, the adoption of wind, solar, and battery storage devices, which are power electronics-interfaced resources, continues to increase worldwide \cite{ExecutiveSummaryRenewables}. However, the physical principals determining the behavior of inverter-based resources (IBRs) are fundamentally different from that of conventional, synchronous machine-interfaced resources. As IBRs supplant synchronous generators (SGs), the dominant system dynamics shift from electromechanical modes to electromagnetic modes \cite{kenyonValidationMauiPSCAD2021}. Thus, this transition from SGs to IBRs warrants new analytical models to best explain the underlying physical phenomena consistent with the nature of emerging power systems.  

Analytical models are the building blocks of algorithms and the core engines behind all existing scientific and commercial simulation tools used in research and industry to predict the system behavior. However, there is always a trade-off between the granularity of results vs. the scale of system that can be simulated with reasonable computational effort. 
A review of the literature suggests that two general directions have emerged in modeling the dynamics of power systems. The first direction focuses on different formulations, such as sinusoid waveform vs. phasor models. Sinusoid waveform models provide high-fidelity simulations of power systems by explicitly modeling all three phases and controllers, supporting the use of high-order, highly detailed device models, and having time step capabilities in the microsecond range \cite{kenyonValidationMauiPSCAD2021}. However, solving waveform models, which is the engine in software packages such as Power System Computer Aided Design (PSCAD), remains impractical for use in large-scale systems due to their computational burden \cite{kenyonValidationMauiPSCAD2021}. 
Alternatively, the use of phasor models, which are reduced-order models, utilize simplified device and network models and consequently their computational tractability is far superior to waveform models. However, while phasor models are attractive due to their  computational tractability,  they are less accurate than full-order models \cite{kenyonValidationMauiPSCAD2021}. Positive sequence RMS (Root Mean Square) simulation tools, such as Power Systems Simulator for Engineers (PSSE) and PowerWorld, are examples of reduced-order models in the sense that they reduce a three-phase explicit model to a positive sequence model. However, these tools, which were developed based on the slow electromechanical dynamics of SGs are ill suited for modeling IBRs due to their large time steps (about 4 ms), and inability to model internal controllers \cite{Zhu2018ModelingIR}. While both the full-order models in EMT tools and the reduced-order models in positive sequence tools have clear shortcomings, there are also other methods of model reduction. 

Order reduction can be achieved by eliminating non-dominant modes in the dynamics of variables of interest, and can be performed through various techniques, e.g., perturbation theory \cite{simmondsFirstLookPerturbation1998}, integral manifolds \cite{KokotovicIntegralmanifold}, or simply through simplified physics-based reduction methods \cite{chanDynamicEquivalentsAverage1972}. Almost universally, existing analytically reduced models for power systems are focused on system frequency (SFR) models, while developing such models for nodal voltage dynamics has not been a focus of the literature \cite{chanDynamicEquivalentsAverage1972,
	AndersonALowOrderSystemFrequency,
	aikAGeneralOrderSystemFrequency,
	ZHANGFrequencyConstrainedUnitCommitment,
	egidoMaximumFrequencyDeviationCalculation,
	liuAnAnalyticalModelForFrequency,
	yangSimplifiedPredictionModel,
	dongUnifiedAnalyticalMethod2023,
	wangThreemachineEquivalentSystem2022}. SFR models that aggregate the primary frequency response of a system take the order-reduction process to an extreme by omitting network effects and neglecting all variance in frequency response across a system \cite{AndersonALowOrderSystemFrequency}. SFR models have been widely used as a less computationally burdensome alternative to time domain simulation for the estimation of the frequency response of a system after a disturbance \cite{chanDynamicEquivalentsAverage1972,
	AndersonALowOrderSystemFrequency,
	aikAGeneralOrderSystemFrequency,
	ZHANGFrequencyConstrainedUnitCommitment,
	egidoMaximumFrequencyDeviationCalculation,
	liuAnAnalyticalModelForFrequency,
	yangSimplifiedPredictionModel,
	dongUnifiedAnalyticalMethod2023,
	wangThreemachineEquivalentSystem2022}. They yield reasonable approximations of RoCoF and nadir frequency for SGs, which can then be formulated into constraints on frequency stability or under-frequency load-shedding in power systems optimization problems \cite{aikAGeneralOrderSystemFrequency,ZHANGFrequencyConstrainedUnitCommitment}. 
In \cite{chanDynamicEquivalentsAverage1972} and \cite{AndersonALowOrderSystemFrequency}, lower-order SFR models are presented, but the SGs exhibit little heterogeneity in terms of governor and prime mover type. These models were intended to provide general information about a system's frequency response post-disturbance \cite{AndersonALowOrderSystemFrequency}, and to formulate margins that would ensure the minimum frequency following a disturbance would not be less than the load-shedding frequency threshold \cite{chanDynamicEquivalentsAverage1972}.
An open-loop model to calculate the maximum frequency deviation in a small isolated power system is proposed in \cite{egidoMaximumFrequencyDeviationCalculation}. This model allows for SGs with different governor, prime mover, and turbine characteristics, unlike the SFR model of \cite{AndersonALowOrderSystemFrequency}. In \cite{liuAnAnalyticalModelForFrequency}, a parabolic frequency deviation serves as the input to an open-loop SFR model to better predict system nadir.  
However, all of these models presume a SG-dominated system and most neglect the network dynamics and the spatial-temporal characteristics of frequency response. In all cases, reactive-power/voltage (Q-V) dynamics are completely disregarded. 

IBR-rich systems typically feature power electronic devices with two prominent inverter control paradigms: grid-following inverters (GFLs) or grid-forming inverters (GFMs). Both devices provide power, but only GFMs construct frequency and regulate voltage independently. GFLs with grid support functionality, i.e. volt-var and volt-watt functions, can also manipulate frequency and voltage response through the regulation of active and reactive power, but they required a well-defined voltage waveform at their point of interconnection and they cannot independently form frequency \cite{linResearchRoadmapGridForming}. For these reasons, we focus on GFMs in this work. With regards to the frequency response of inverters, the SFR model of \cite{yangSimplifiedPredictionModel} is intended to model systems with some amount of renewable generation, though all renewable generators are modeled as doubly fed induction generators with droop control. In \cite{dongUnifiedAnalyticalMethod2023}, a nadir prediction model is proposed that accounts for three different types of fast frequency response provided by IBRs. However, the authors neglect the fast dynamics of IBRs and do not distinguish between GFL and GFM. 
The model in \cite{wangThreemachineEquivalentSystem2022} omits inverter dynamics but makes the unique contribution of providing a closed-form solution of frequency response for a three SG system. In this work, a simplified network model is used to account for inter-machine frequency oscillations. However, application of \cite{wangThreemachineEquivalentSystem2022} to large, multi-generator systems requires the aggregation of generators into three equivalent SGs. Importantly, the model presumes a flat voltage profile of $1\text{pu}$ across the system. The review of the literature on voltage models also reveals that existing reduced order models of GFM networks that account for voltage dynamics require the division of a network into an aggregated ``external area" and a ``study area" \cite{duModelReductionforInverter} and require the inclusion of currents as state variables \cite{ajalaModelReductionandDynamicAggregation}. Additionally, to the best of our knowledge, no low-order model exists to specifically compute nodal voltage and frequency for \textit{all GFM nodes} in large-scale power systems.

This work advances the field by using physics-based reduction methods to develop low-order, accurate models for frequency \textit{and voltage dynamics} in all-inverter systems. GFM devices are modeled explicitly, while GFL inverters are modeled as negative load. 
The models are accurate and scalable, effectively eliminating the need for EMT simulations for a variety of applications. To this end, we first demonstrate the critical role of reactive power losses post-disturbance and its effect on voltage response and present novel methods for calculating post-disturbance reactive power participation factors for each GFM in a system. We then derive our proposed analytical models for frequency and voltage response for all-inverter systems. We utilize matrix exponentials to dramatically accelerate computation time. Using EMT simulation, we demonstrate the significant advantages that our models present with regards to solution speed and scalability, while preserving high accuracy. For these reasons, the models lend themselves well to direct application within industry and research environments. The ability to accurately and efficiently simulate transients in all-inverter systems sets the basis for probabilistic analysis of frequency and voltage stability as well as security-constrained economic dispatch and real-time dynamic security assessment, among other potential future applications.

\section{The Need For New Analytical Models}\label{problemstatement}
In this section, we argue that the need for new analytical models of system dynamics is threefold. 
First, existing frequency response models are for 100\% SG systems, or systems that are dominated by SGs, and are therefore intrinsically inappropriate for modeling all-inverter systems. 
Secondly, existing system frequency models neglect the heterogeneity of frequency response and the impacts of the transmission network, which we will demonstrate is particularly unsuitable for low-inertia, all-inverter systems. 
Finally, while existing models capture P-$\delta$ dynamics reasonably well, they assume Q-$\delta$ dynamics are insignificant, and as a result, voltage profiles are assumed to be 1 pu, which is only realistic for a network with large SGs. Hence, there is need for an analytical model that reflect this non-negligible coupling between active power and voltage, especially in all-inverter systems where these dynamics are even more critical \cite{sajadiPlaneWaveDynamic}.
In what follows, we discuss each of the above-mentioned issues in detail and present evidence in support of our arguments.

\subsection{Frequency response: SG vs. GFM}
\begin{figure}[htbp]
	\centering
	\includegraphics[width=0.88\columnwidth,trim={0 0 0 0},clip]{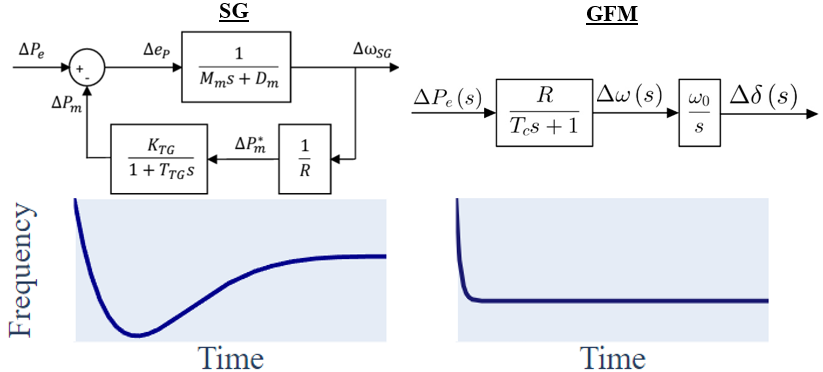}
	\caption{Frequency response block diagram and response shape for: 1) SGs (left) and 2) GFMs (right) \cite{sajadiSynchronizationElectricPower2022}.}
	\label{fig:SG_GFM}
\end{figure}

The frequency response of all inverter systems is markedly different from that of conventional systems, where SGs provide frequency response. This difference arises from the fundamentally different way power and frequency are generated by GFMs vs. SGs. The frequency dynamics in SGs are initially governed by the mechanical momentum of the turbine, which determines the rate at which frequency deviates. This ``inertial phase" is followed by primary frequency control, where the action of the turbine governor changes the generator's active power output to arrest the frequency deviation \cite{machowskiPowerSystemDynamics2020}. The rate of change of frequency is entirely dependent on the generator momentum \cite{sajadiPlaneWaveDynamic}, commonly expressed in seconds and referred to as the inertia constant. In primary frequency control, active power output changes as a function of the speed-droop coefficient, which is typically 5\% \cite{machowskiPowerSystemDynamics2020}. Accordingly, the amount of momentum in the system and the governor response of the online generators collectively determine the frequency nadir, which is the point on the frequency response curve where frequency is lowest \cite{liuAnAnalyticalModelForFrequency}. The frequency response mechanism of a generic SG is shown in Fig. \ref{fig:SG_GFM}. 

While Newtonian physics guides frequency deviations in SGs, forming a second-order dynamical system, the frequency response of inverters is purely a function of their control algorithms; thus, as shown in \cite{kenyonInteractivePower}, it presents a first-order dynamical system on the same timescales. There are three prominent control regimes for GFMs including linear droop, virtual synchronous machines, and virtual oscillator control \cite{kenyonInteractivePower}. In this paper, we focus on multi-loop droop-control GFM (Fig. \ref{fig:SG_GFM}), due to their popularity and markedly different response shape as compared to SGs. In a droop control GFM, frequency is adjusted according to the device's droop gain constant, which can be changed and is not intrinsic to the device \cite{kenyonInteractivePower}.  
Due to their very small inertia constant, lower nadirs are often observed in power systems with high shares of inverters
\cite{linResearchRoadmapGridForming}, which is undesirable due to the legacy protection systems \cite{osti_1769842}. High RoCoF and low nadir values could unintentionally activate protection mechanisms, resulting in load-shedding. This was exactly the case in Britain on August 9\textsuperscript{th}, 2019, when the Low Frequency Demand Disconnection (LFDD) protection mechanism was activated, tripping off massive amounts of distributed generation \cite{GBPowerSystem2020}. Therefore, it is important to develop computationally efficient analytical models that adequately capture frequency response in large systems with high shares of GFMs.

\subsection{Network frequency heterogeneity}

\begin{figure}[h]
	\centering
	\includegraphics[width=.8
	\columnwidth,trim={0 0 0 0},clip]{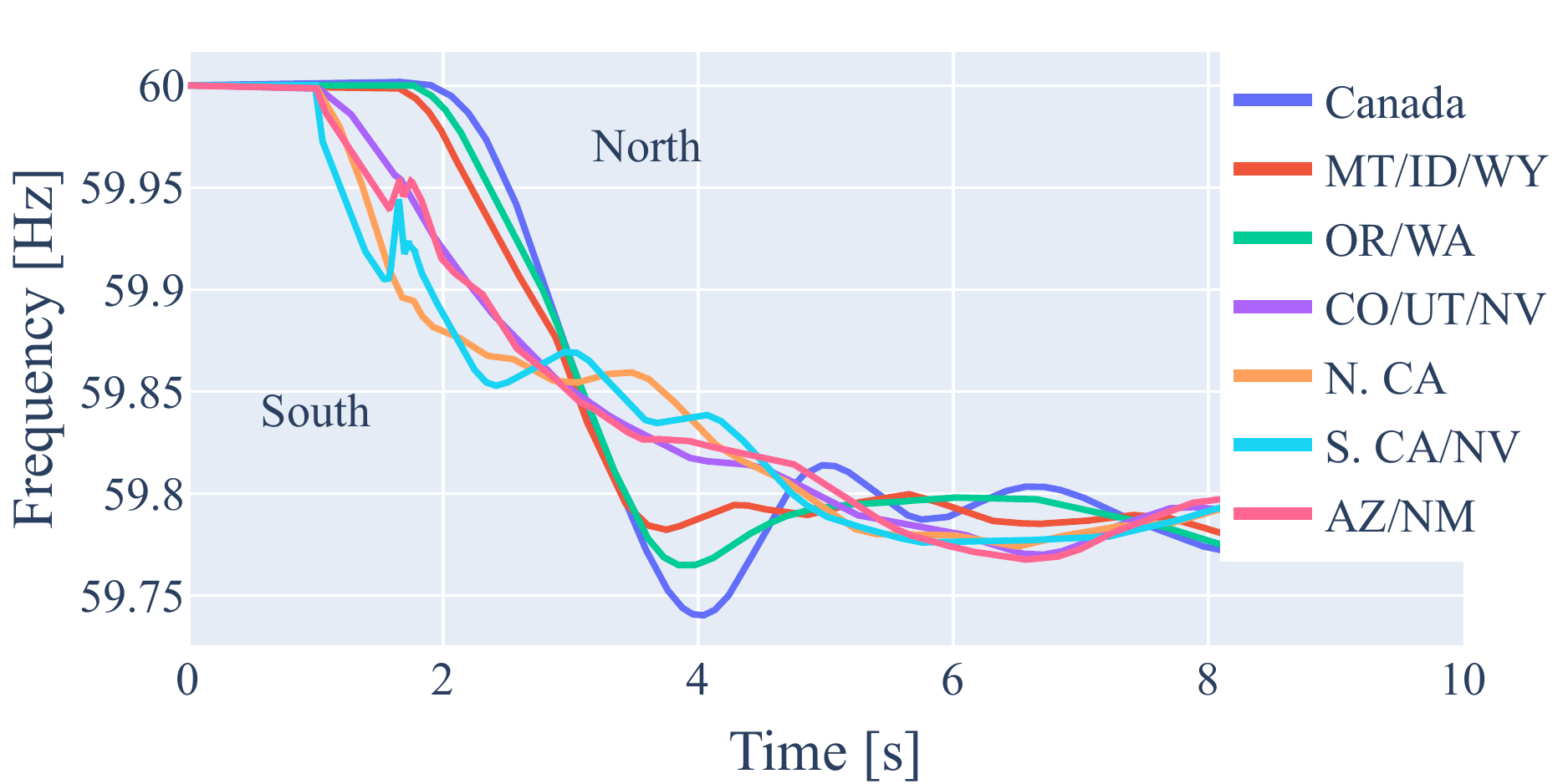}
	\caption{PowerWorld simulation results of the WECC system following the outage of the Palo Verde Nuclear Generating Station, the system dimensioning contingency.}
	\label{fig:WECC}
\end{figure}

Many system frequency models are founded upon the operating principles of SGs and on the assumption of frequency being homogeneous across the network. Under this assumption, a system frequency model can be formulated by conveniently aggregating the parameters of SGs and modeling them as a single representative generator \cite{chanDynamicEquivalentsAverage1972}. In such a model, there is no need to model how the generators in the system synchronize with each other and the impact of network topology is neglected. In reality, frequency is not homogeneous across a power network. Rather, disturbance propagation -- and therefore frequency response -- is closely related to network topology. More specifically, disturbances are more severe in areas supporting the Fiedler eigenvalue of the network Laplacian \cite{pagnierInertiaLocationSlow2019}. Therefore, by aggregating generator parameters to approximate a ``system frequency response," one is ignoring the impact of network connectivity and systematically underestimating the maximum RoCoF and frequency deviation. To illustrate this effect, we simulate the outage of the Palo Verde Nuclear Generating Station in the WECC system, and the results are provided in Fig. \ref{fig:WECC}. 

In Fig. \ref{fig:WECC}, it is evident that even in this primarily SG network the frequency dynamics are not homogeneous, with varying values of RoCoF and nadir in each region. This suggests that neglecting the spatial-temporal characteristics of frequency response in a system yields inaccurate estimations of key metrics, particularly for large systems. Heterogeneity in frequency response could be exacerbated by higher shares of IBRs and a lack of inertia, leading to the simultaneous occurrence of frequency, voltage, and network dynamics \cite{sajadiPlaneWaveDynamic}. Additionally, there is clearly a time delay between when significant deviation from nominal frequency begins to occur in the southern region versus in the northern region. Time delays between frequency response trajectories is another example of detail that is lost when representing the frequency response of a system with a single trajectory.

\subsection{Voltage response and reactive power flow}\label{sec:vReponseModels}
While active power is the part of apparent power that does the work, reactive power is vitally important for preventing the demagnetization of transmission lines and sustaining voltages during transients \cite{sajadiPlaneWaveDynamic}. Without adequate reactive power provision, low voltages may occur and contribute to a blackout \cite{north2004technical}.
The changing nature of frequency in IBR-dominated power systems also has ramifications in terms of voltage dynamics. While active and reactive power are closely related, frequency and voltage are treated independently in the conventional wisdom of power system analysis \cite{machowskiPowerSystemDynamics2020}. This is because in a conventional power system dominated by SGs, automatic voltage regulators (AVRs) locally control the terminal voltage of SGs by adjustment of their field currents \cite{machowskiPowerSystemDynamics2020}. Because AVR dynamics are electromagnetic in nature, voltage regulation happens at a much faster timescale than frequency response, which is an electromechanical process in a SG dominated environment. Therefore, frequency and voltage dynamics are evaluated separately \cite{sajadiPlaneWaveDynamic}. 
Due to the long-standing and ubiquitous treatment of voltage and frequency as decoupled, knowledge about the limits of the validity of the decoupling approach is lacking. At the same time, there is more need than ever to build upon our understanding of P-V and Q-$\delta$ dynamics due to new challenges posed by the introduction of IBRs. Unlike SGs, where frequency and voltage dynamics occur on different timescales, the frequency and voltage response of IBRs are both electromagnetic in nature \cite{milanoFoundationsandChallenges}. The contemporaneousness of these dynamics may result in unexpected coupling dynamics and may lead to instabilities and low security \cite{milanoFoundationsandChallenges}. 
Also distinguishing IBR dynamics from SG dynamics are constraints related to the silicon-based switches of power electronics. To avoid damaging these switches, IBRs limit their active and reactive power injections, whereas the power output of SGs is not bounded to the same degree; they can temporarily deviate from the excitation setpoint and supply a transient response which greatly exceeds the nominal rating of the machine \cite{KenyonAGrid-FormingPair}. In other words, while SGs are capable of providing generous active and reactive power during transients to support voltage and frequency, IBRs must more carefully provide power as to not violate current limits. As a result, the active and reactive power outputs of an IBR are closely related \cite{sajadiPlaneWaveDynamic}.
Here, we demonstrate the increased coupling in emergent power networks with all GFMs and the need to include both active and reactive power distribution factors.

\begin{figure}[h]
	\centering
	\includegraphics[width=.8\linewidth]{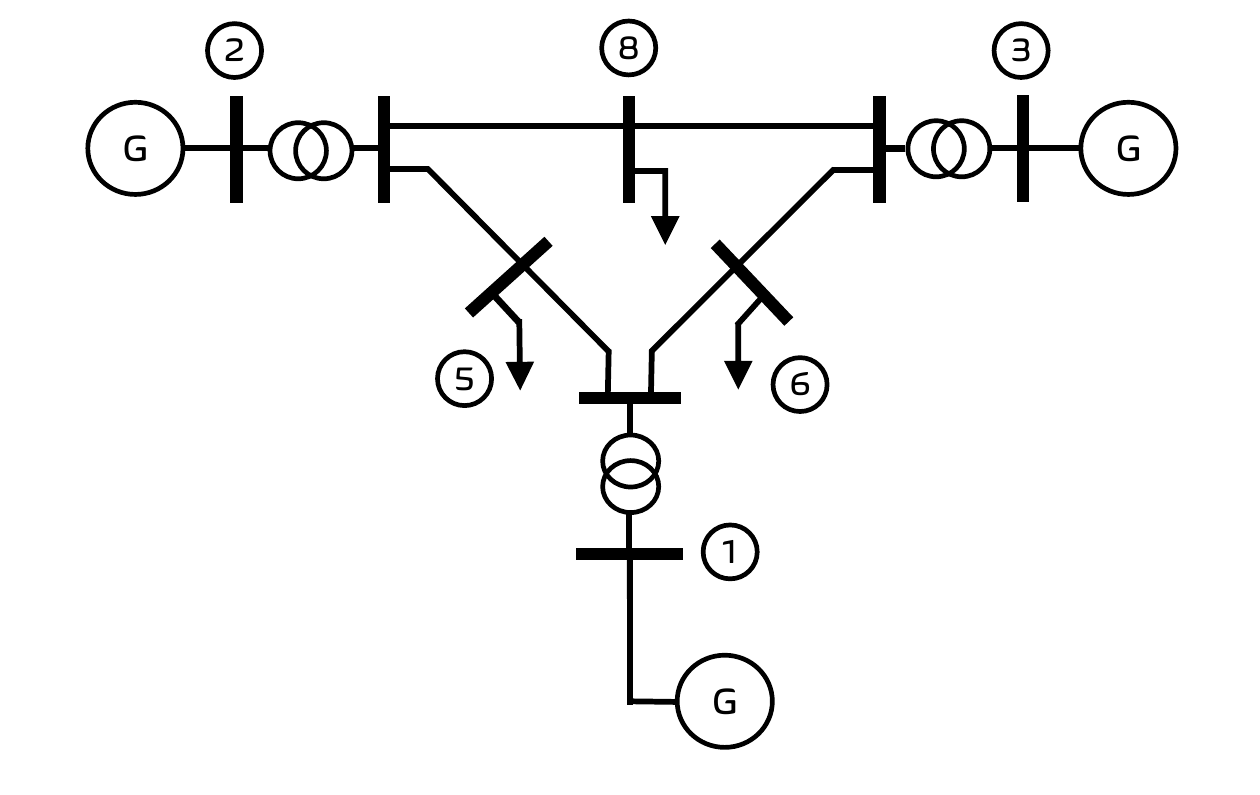}
	\caption{One-line diagram of the WSCC 9-bus system.}
	\label{fig:bus9_system}
\end{figure}

Estimating Q-V dynamics post-disturbance is difficult due to the nonlinear nature of reactive power flow \cite{gentileOnReactivePowerFlow}. Interestingly, active power participation factors can still be reasonably estimated while neglecting Q-V dynamics, and \cite{chanDynamicEquivalentsAverage1972,
	AndersonALowOrderSystemFrequency,
	aikAGeneralOrderSystemFrequency,
	ZHANGFrequencyConstrainedUnitCommitment,
	egidoMaximumFrequencyDeviationCalculation,
	liuAnAnalyticalModelForFrequency,
	yangSimplifiedPredictionModel,
	dongUnifiedAnalyticalMethod2023,
	wangThreemachineEquivalentSystem2022
} leverage this fact. To demonstrate this point, we simulate two different load steps in the WSCC 9-bus system, depicted in Fig. \ref{fig:bus9_system}, and calculate the active and reactive power participation factors of each generator. The system has 3 GFM inverters and 3 constant power loads. Assuming generators have adequate headroom, we assert that the active and reactive participation factors of each generator can be estimated by solving the nonlinear power flow equations following a disturbance. The participation factors are equal to the change in power injection pre- and post-disturbance at each generation node. The calculation differs from the standard AC power flow because, instead of buses being categorized as $V\theta$ (slack), $PV$, and $PQ$ types, they are only $V\theta$ and $PQ$ type. Buses with GFMs are treated as $V\theta$ buses immediately after a disturbance and before the inverter controls have time to adjust their voltage angle and magnitude setpoints in response to changes in tie-in active and reactive power.

First, an active power load step of $31.5\text{ MW}$ is applied to Bus 6 and the active power participation factors are calculated. Then, a load step of $31.5\text{ MW}$ plus $9\text{ MVar}$ was applied to Bus 6, and again the active power participation factors are calculated. Note that $31.5\text{ MW}$ is $10\%$ of the active power load and $9\text{ MVar}$ is 10\% of the reactive power load. The results, which are shown in Table \ref{tab:postD_P}, indicate that even when a 10\% reactive power load step is applied, the active power participation factors do not differ greatly.

\begin{table}[htbp]
	\centering
	\caption{Percent difference change in active power injection at generation buses, $\Delta  P_G$, between: 1) an active power load step, 2) an active and reactive power load step.}    
	\begin{tabular}{c|ccc}
		& Bus 1 & Bus 2 & Bus 3\\\hline
		\% Difference $\Delta P_G$ & 0.4612 & 0.9597 & 0.6341 \\
	\end{tabular}
	\label{tab:postD_P}
\end{table}

The picture is very different for changes in reactive power injection following a complex power disturbance. In Table \ref{tab:changingDeltaQ} the results of participation factor calculations following seven disturbances are shown. In each of these simulations, a reactive power disturbance of $9\text{ MVAR}$ was applied to Bus 6. An active power step ranging from $0\text{ MW}$ to $30\text{ MW}$ was also applied in each simulation. The baseline $\Delta Q_G$ values were calculated following a 9 MVAR load step. Even though the magnitude of the reactive power load step was the same for each simulation, the reactive power injection at each generation node increases with the magnitude of the active power disturbance. For example, $\Delta Q_G$ at Bus 1 was 12.13\% greater than the baseline value following the addition of a 5 MW load step. To explain this behavior, recall the equation for reactive power at the sending end of a transmission line (\ref{QR}) \cite{machowskiPowerSystemDynamics2020}.

\begin{table}[htbp]
	\centering
	\caption{Percent difference in reactive power injection at generation buses, $\Delta  Q_G$, after complex load steps with increasing active power disturbance magnitudes. }
	\begin{tabular}{c|ccccccc}
		\% Error in & \multicolumn{7}{c}{Active Power Disturbance Magnitude [MW]}\\
		$\Delta Q_G$ &0& 5 & 10 & 15&20&25&30\\\hline
		Bus 1 & 0 & 12.13 & 24.77 & 37.92 & 51.57 & 65.75 & 80.44\\
		Bus 2 & 0 & 16.75 & 34.01 & 51.78 &  70.06 &  88.87 & 108.21\\
		Bus 3 & 0 & 13.26 & 27.02 & 41.29 & 56.08 & 71.38 & 87.21\\
	\end{tabular}
	\label{tab:changingDeltaQ}
\end{table}

\begin{equation}\label{QR}
	Q_S = \Im\{V_S I_S^*\} \approx \dfrac{V_S}{X}(V_S  - V_R\cos\delta_{SR} )
\end{equation}

\begin{figure}
	\centering
	\begin{subfigure}[b]{0.18\textwidth}
		\includegraphics[width=\textwidth]{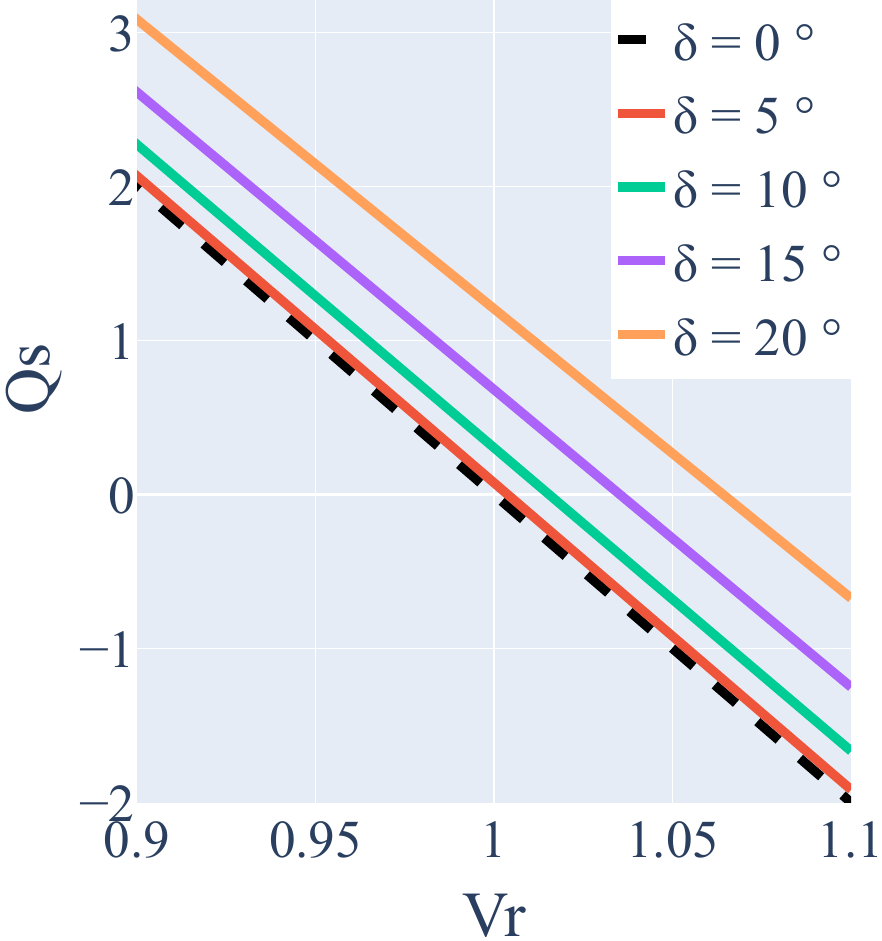}
		\caption{$Q_S$ vs. $V_R$ for different values of $\delta_{SR}$.}
		\label{fig:3}
	\end{subfigure}
	\hfill
	\begin{subfigure}[b]{0.27\textwidth}
		\includegraphics[width=\textwidth]{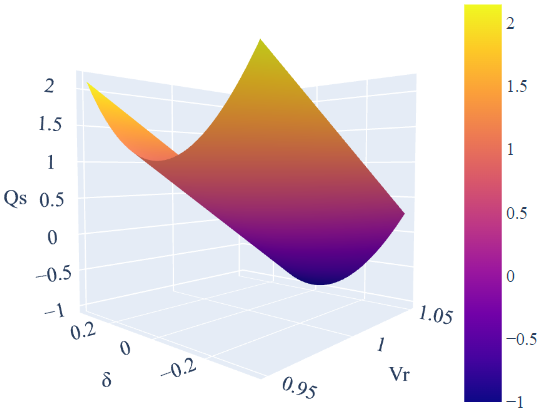}
		\caption{Sending-end reactive power as $V_R$ and $\delta_{SR}$ change.}
		\label{fig:2}
	\end{subfigure}
	
	\caption{Sending-end reactive power with respect to $\delta_{SR}$ and receiving-end voltage. $V_S = 1.0$ and $X = 0.05$.}
	\label{fig:QVdelta}
\end{figure}

Usually, $\delta_{SR}$ is assumed to be small and $\cos\delta_{SR} \approx 1$. This yields the trace corresponding to $\delta = 0 \degree$ in Fig. \ref{fig:3} \cite{machowskiPowerSystemDynamics2020}, which is shown as a dashed line. After a disturbance, however, the angle across a transmission line $\delta_{SR}$ may increase, impacting reactive power injections across the system. If we choose not to approximate $\cos\delta_{SR}\approx 1$, we have a more complete picture of reactive power dynamics, as shown in Fig. \ref{fig:QVdelta}. From this figure, we see there is significant curvature along the $\delta_{SR}$ axis, which illustrates there is a non-negligible relationship between voltage angle and reactive power. 
While Q-$\delta$ coupling is ignored in some studies \cite{ilicRedistributionofreactivepower}, other works have attempted to account for this behavior. In \cite{singhImprovedvoltageandreactivepowerdistributionfactors} and \cite{ruizPostContingencyVoltage}, Q-$\delta$ dynamics are acknowledged but it is assumed that the slack bus alone changes its power injection to account for generation/load active power imbalances. In reality, all online generators will respond to the changes in states seen at their point of interconnection and therefore this assumption should not be made when attempting to estimate post-disturbance reactive power flow. In this paper, we improve upon previous efforts to estimate post-disturbance reactive power injections by distributing increased active power imbalances more realistically.

In summary, we established that the frequency transients across a power network could vary significantly due to individual generator properties, the fast nature of IBR frequency response mechanisms, and the impact of the transmission network. This frequency heterogeneity is especially pronounced in power networks with high shares of inverters. Moreover, to the author's knowledge, no low-order voltage response model has been developed for either SG-dominated or IBR-dominated systems. This constitutes a massive gap in our ability to estimate post-disturbance dynamics. There is an even greater need for models of Q-V dynamics post-disturbance seeing that the increased deployment of IBRs can give rise to weak grid conditions and compromise voltage stability \cite{dozeinSystemStrengthWeak2018}. These gaps in knowledge serve as the motivation to develop our frequency and voltage models for all-inverter systems, which will be discussed in the next section. 

\section{Proposed Models}\label{models}
In Section \ref{sec:vReponseModels} we demonstrate how $Q$-$V$ dynamics have a limited effect on active power participation factors and frequency. Conversely, changes in voltage angle have a non-negligible effect on the reactive power participation of each generator. 
Therefore, one can estimate frequency well without accounting for reactive power, but cannot estimate reactive power well without accounting for changes in phase angle. 
Given this information, we propose a frequency response model for all-inverter power systems that assumes decoupled dynamics, and a voltage response model that accounts for Q-$\delta$ dynamics. These models can be solved independently.
\subsection{Frequency Response Model}
The reduced-order block diagram representing the frequency response of a multi-loop droop GFM is shown in Fig. \ref{fig:SG_GFM}. The power error is detected through a low pass filter, creating a power error signal. This error signal serves as the input to the droop controller \cite{sajadiSynchronizationElectricPower2022}. $T_{c}$ is the reciprocal of the cut-off frequency in $\text{Hz}$ and $R$ is the droop gain. The corresponding linearized dynamics for the $i$th machine are given by system~\eqref{sys0}:
\begin{align}\label{sys0}
    \begin{cases}
        \Delta \dot{\delta}_i= \omega_{0} \Delta \omega_i, \\
        \Delta \dot{\omega}_i = \frac{1}{T_{ci}}(\alpha_i R_i \Delta P_{ei} -\Delta \omega_i), 
    \end{cases}
\end{align}
where $\delta_i$ is the voltage angle, $\omega_i$ is the frequency, and $\omega_0=2 \pi f_0$, where $f_0$ is the nominal network frequency, e.g. $60 \text{ Hz}$ in the United States. The parameter $\alpha_i$ is equal to the system base divided by

Real and reactive nodal power injections are nonlinear functions of voltage magnitude and angle. These functions are typically linearized around a particular operating point \cite{machowskiPowerSystemDynamics2020}. Our frequency response model utilizes DC power flow approximations. \begin{comment}These approximations yield the following network equations for active power.
%
\begin{equation}\label{P_dc}
 P_{k} = \sum_{j=1}^{N}(B_{kj}(\delta_{k}-\delta_{j}))
\end{equation}
\end{comment}
In the DC formulation, reactive power flow is disregarded because it is smaller than active power flow. The resulting linearized equation for power is (\ref{dcpf_inc}), which can be written in terms of power injections at generation and load nodes as seen in (\ref{networkeqn}). Grid-following inverters without grid support functionality may be treated as load nodes. 
\begin{equation}\label{dcpf_inc}
\Delta P = \bB \Delta \delta
\end{equation}
\begin{equation}\label{networkeqn}
 \begin{bmatrix} \Delta P_{G} \\ \Delta P_{L}\end{bmatrix} =  \begin{bmatrix} \bB_{GG} & \bB_{GL} \\ \bB_{LG} & \bB_{LL} \end{bmatrix} \begin{bmatrix} \Delta \delta_{G} \\ \Delta \delta_{L}\end{bmatrix}
\end{equation}
The network is topologically reduced through the elimination of non-generator nodes \cite{machowskiPowerSystemDynamics2020}, as follows:
\begin{equation}\label{delta_p_gen}
\begin{split}
\Delta P_{G} &= \underbrace{{\bB_r} \Delta \delta_{G}}_{\Delta P_{os}} + \underbrace{\bB_L\Delta P_L}_{\Delta P_{d}}  \\
\end{split}
\end{equation}
where
$\bB_{r}= \bB_{GG}-\bB_{GL}\bB_{LL}^{-1}\bB_{LG}$ and $ 
\bB_L= \bB_{GL}\bB_{LL}^{-1} $.
From (\ref{delta_p_gen}) we see that the change in power at a generation node is a function of the differences in angles between generation nodes and the change in power injections at the non-generator nodes \cite{wangThreemachineEquivalentSystem2022}. We refer to the first term of (\ref{delta_p_gen}) as the inter-machine oscillation of power. The second term, which is denoted as $\Delta P_d$ is interpreted as the change in power injections at generation nodes due to a disturbance. $\Delta P_d$ is a vector with a number of elements equal to the number of generators. $\Delta P_{L}$ is the vector of load changes. For instance, consider a system with three load buses and a load step at load bus 3. The third element of $\Delta P_{L}$ is then $P_{step}*K_{l}$, where $K_{l}$ is the approximated marginal loss constant. Average transmission and distribution losses are about 2-5\%, and therefore we recommend choosing 1.05 as the value of $K_{l}$ to avoid overestimating nadir and settling frequency \cite{FrequentlyAskedQuestionsa}. The approximation of a linear relationship between active power transfer and line losses is consistent with the concept of marginal loss factors \cite{eldridgeMarginalLossCalculations2016}. The higher the value of $K_l$, the more conservative the estimate of nadir and settling frequency. It should be noted that non-generator nodes are assumed to be static and, therefore, this model is limited in capturing load dynamics. The modeling of dynamic loads, including grid-following inverters with frequency support functionality, necessitates the preservation of the relevant nodes during Kron reduction which are not considered in this work. For a system with $n$ droop control GFMs, we define
%\begin{align*}
\begin{equation}
    \Delta \delta \triangleq \left[\Delta \delta_{1,n} \; \hdots \; \Delta \delta_{n-1,n}\right]^\mathrm{T}  
\end{equation}
   % [2mm]
   \begin{equation}
    \Delta \omega \triangleq \left[ \Delta \omega_1 \; \hdots \; \Delta \omega_n\right]^\mathrm{T} 
    \end{equation}
%\end{align*}
where $\Delta \delta_{i,n}$ denotes the relative voltage angle $\Delta \delta_{i} -  \Delta \delta_{n}$. Relative voltage angles are used so that synchronizing dynamics may be represented \cite{machowskiPowerSystemDynamics2020}. Then, the system-wide dynamics are given by the $(2n-1)$-dimensional system of ODEs in \eqref{state_space}. 
\begin{equation}\label{state_space}
\frac{\mathrm{d}}{\mathrm{d}t}
 \begin{bmatrix} \Delta \delta  \\ \Delta \omega \end{bmatrix} =  \bA_f\begin{bmatrix} \Delta \delta\\ \Delta \omega\end{bmatrix}  + \bB_f \Delta P_d, 
 \end{equation}
 where 
 \begin{equation}
     \bA_f = 
     \begin{bNiceArray}{c|c}[margin,baseline=c]
     \bzero & \omega_0\bone_{-\bone} \\
     \hline 
     \tbB & \bT 
     \end{bNiceArray} \quad \text{ and } \quad 
     \bB_f = 
     \begin{bNiceArray}{c}[margin]
         \bzero \\
         \hline
         \bgam
     \end{bNiceArray} 
 \end{equation}
     with 
     \begin{equation}
         \bone_{-\bone} = 
         \begin{bmatrix}
             1 & 0 & \hdots & 0 & -1 \\
             0 & 1 & \hdots & 0 & -1 \\
             \vdots & \vdots & \ddots & \vdots & \vdots\\
             0 & 0 & \hdots & 1 & -1
         \end{bmatrix} \in \R^{(n-1) \times n}, 
     \end{equation}
     \begin{equation}
         \tbB = -\diag\left(\frac{\alpha_i R_i}{T_{ci}}\right)\tbB_r, \quad \bT = -\diag\left(\frac{1}{T_{ci}}\right), 
     \end{equation}
     and 
     \begin{equation}
         \bgam = \diag\left(\frac{\alpha_i R_i}{T_{ci}}\right). 
     \end{equation}
     $\tbB_r$ denotes the matrix $\bB_r$ with the $n$th column removed. 
 We refer to the proposed frequency response model as the Low-Inertia Frequency Evolution (LIFE) model. The model is intended for modeling the dynamic behavior of a system following a small to moderate power imbalance. Scenarios where very large voltage angle differences across transmission lines ($>30$°) are seen or where large changes in voltage magnitude are observed, i.e. during faults and restoration processes, are not captured by the model. 
\subsection{Voltage Response Model}
\begin{figure}[htbp]
	\centering \includegraphics[width=.15\columnwidth,trim={0 0 0 0},clip,angle=90]{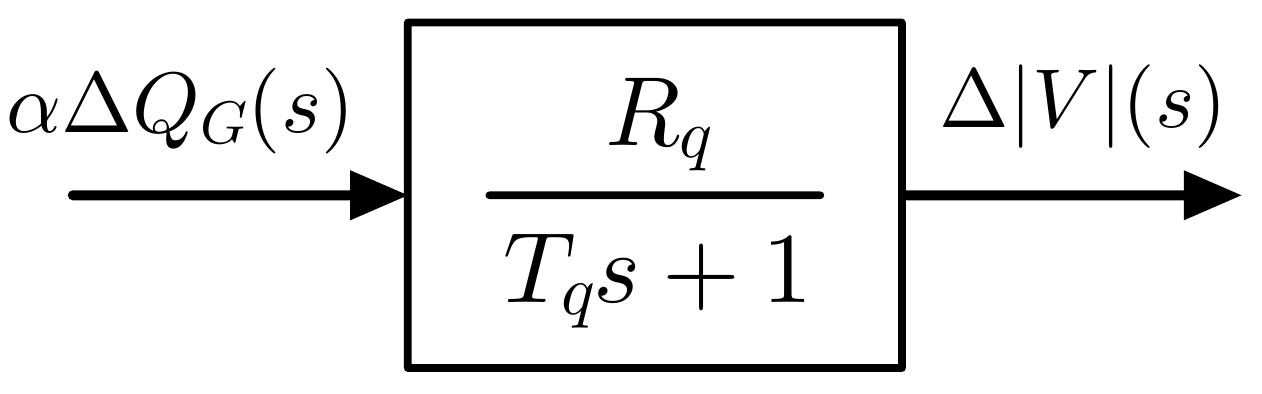}
	\caption{Low-order model of reactive droop.}
	\label{fig:Q_droop}
\end{figure}

The LIFE model made the standard assumption of decoupled active and reactive power flow. Here, we propose a model of voltage response that accounts for post-disturbance Q-$\delta$ coupling and can be thought of as the counterpart of our frequency response model. The reduced-order block diagram representing the reactive power droop of a multi-loop GFM is given in Fig. \ref{fig:Q_droop}. A detailed discussion on the validity of this model and on determining an appropriate time constant, $T_q$, is included in the companion Supplemental Note of this paper. $R_{q}$ is the reactive power droop value.

Given that reactive power cannot travel long distances and voltage magnitudes do not synchronize across a network, the open loop model of Fig. \ref{fig:Q_droop} is a reasonable method for estimating voltage response locally \cite{machowskiPowerSystemDynamics2020}. The model must be provided with a good approximation of the change in reactive power injection at the generator nodes, $\Delta Q_G$. The equation for instantaneous reactive power injection is (\ref{Qi}).
\begin{equation}\label{Qi}
	Q_i = -\sum^n_{j=1}B_{ij}V_iV_j\cos\theta_{ij} + \sum^n_{j=1}G_{ij}V_iV_j\sin\theta_{ij}
\end{equation}
Assuming the reactive power set points of all GFMs are known, the change in reactive power injection due to the disturbance can be described as (\ref{deltaQG}).
\begin{equation}\label{deltaQG}
	\Delta Q_G = Q_G(V_{post},\theta_{post}) - Q_{G0}
\end{equation}
where $V_{post}$ and $\theta_{post}$ are the post-disturbance angles and voltages, and $Q_{G0}$ is the inverter Q set point. We propose the following steps for estimating $Q_G(V_{post},\theta_{post})$ and $\Delta Q_G$.
\begin{enumerate}
	\item Estimate each new GFM active power injection using $P_{Gi}(V_{post},\theta_{post}) \approx \Delta P_{di} + P_{Gi0}$ where $\Delta P_{di}$ is calculated in (\ref{delta_p_gen}).
	\item In the AC power flow formulation, update the active power injections to reflect $P_{Gi}(V_{post},\theta_{post})$, $\Delta P_L$, and $\Delta Q_L$. Then solve AC power flow. 
	\item The updated Q values are approximately equal to $Q_G(V_{post},\theta_{post})$. 
	\item Approximate $\Delta Q_G$ using (\ref{deltaQG}).
\end{enumerate}
The difference between solving for $Q_G(V_{post},\theta_{post})$ as described in Section \ref{sec:vReponseModels} and in the process enumerated above boils down to the assumed known quantities at each generator bus. In a classic AC power flow problem, $P$ and $V$ are the known quantities at the generation buses, while in the proposed formulation of Section \ref{sec:vReponseModels}, $V$ and $\theta$ are the known quantities. However, since P and $\theta$ are closely related, these approaches yield nearly identical results, as shown in Table \ref{tab:deltaQ_methods}. The benefit of using the process  enumerated above is that it allows for the leveraging of AC power flow solvers, which are ubiquitous and effective for large systems.

\begin{table}[htbp]
	\centering
	\caption{Validation of proposed reactive power injection approximation method.}
	\begin{tabular}{c|ccc}
		& Bus 1 & Bus 2 & Bus 3\\\hline
		\% Error of $\Delta Q_G$ & 0.2947 & 0.7333 & 0.2789 \\
	\end{tabular}
	\label{tab:deltaQ_methods}
\end{table}
In a droop-controlled GFM, the quadrature-axis (q-axis) voltage is adjusted according to the droop value $R_q$ and the change in reactive power injection $\Delta Q_G$ \cite{kenyonOpenSourcePSCADGridFollowing2021}. This adjusted q-axis voltage then passes through the voltage controller and current controller before the q-axis and direct-axis (d-axis) voltages are transformed into the phase voltages $V_a$, $V_b$, and $V_c$ via the inverse Park Transformation. These voltages serve as the inputs to the inverter LCL filter \cite{kenyonOpenSourcePSCADGridFollowing2021}. As a result, the change in voltage magnitude at the generator bus is actually several times larger than $\Delta V_G$ as calculated in (\ref{V_deq}). Therefore, to calculate the exact change in voltage magnitude at the generator bus would require modeling these additional controls and including the q-axis and d-axis currents and voltages. This would increase the number of states in the system several times over, which would severely diminish the computational tractability of the voltage response model for large systems and precisely the bottleneck of the present day EMT simulation platforms, and thus defies the original purpose of reduced-order models. This address this limitation, we propose the introduction of the voltage outer-loop-to-grid constant, $K_{v}$, to maintain computational efficiency of the voltage response model. This constant accounts for the aggregation of all coefficients involved, including the equivalent gain factor between the droop-adjusted $V_q$ and the post voltage and current controller $V_q$, and in effect maps the dominant modes that are excited in voltage transients. Depending on the specific GFM inverter model, this constant may vary and the precise value for a specific inverter should be determined by sensitivity testing, either hardware testing in the laboratory environment or system identification analysis in the simulation environment, as described in the companion Supplemental Note of this paper. In our case, after extensive, computational experimentation with industry-grade models for a range of disturbances, we determined values between $4$ and $5$ are most appropriate for the models used in \cite{kenyonOpenSourcePSCADGridFollowing2021}. Further information on the calculation of $K_v$ parameter and a discussion sensitivity testing and mathematical analysis is offered in the Supplemental Note. The change in voltage magnitude at a generator bus is approximated by (\ref{deltaV}), where $V_{G,grid}$ is the voltage at the terminal of the GFM. We subsequently refer to this model of voltage response, which is given in state-space form in (\ref{V_deq}), as the Low-Inertia Voltage Evolution (LIVE) model. In this formulation, $\bB_v$ encodes the settling voltage and $\bA_v$ encodes the pace at which the settling voltage is reached. We note that the proposed voltage model is not intended to simulate dynamics introduced by device malfunctions, i.e. forced oscillations \cite{LesieutreAsystem response}, super-synchronous events, or by AC faults.
\begin{equation}\label{deltaV}
	\Delta V_{G,grid} \approx K_v \cdot \Delta V_G
\end{equation}
\begin{equation}\label{V_deq}
	\begin{bmatrix} \Delta|\dot{V}_{G,grid}|  \end{bmatrix} =  \mathbf{A}_v \begin{bmatrix} \Delta|V_{G,grid}|\end{bmatrix} + \mathbf{B}_v\begin{bmatrix} \Delta Q_{G}\end{bmatrix}
\end{equation}
where
\begin{equation}\label{AvBv}
	\mathbf{A}_v =  -\diag\left(\frac{1}{T_{qi}}\right), \text{ } \mathbf{B}_v = \diag\left(\frac{\alpha_iR_{qi}K_v}{T_{qi}}\right)
\end{equation}

\section{Analytic Solutions and Workflow}
%\subsection{Computational Solver and Acceleration}
%
The proposed models give rise to initial value problems that have analytic solutions; namely, both system \eqref{state_space} and \eqref{V_deq} are of the form
\begin{equation*}
\dot x(t) = \bA x(t) + \bB u(t), \quad x_0 = x(t_0), \quad u_0 = u(t_0),
\end{equation*}
for constant matrices $\bA$ and $\bB$ and known forcing $u(t)$. The solution to such a system is given by 
\begin{equation*}
    x(t) = e^{\bA t}x_0 + \int_{t_0}^t e^{\bA(t - \tau)} B u(\tau) \ \rmd \tau
\end{equation*}
(see \cite{ContTheory}). In both the LIFE and LIVE models, a zero initial condition and constant forcing are assumed, so the solutions simplify to 
\begin{align}
    \begin{bmatrix}
        \Delta \delta \\
        \Delta \omega
    \end{bmatrix}(t) = \int_{t_0}^t e^{\bA_f(t - \tau)}\bB_f\Delta P_d \ \rmd \tau \quad \text{ (LIFE model)}, \label{LIFE soln}\\
     \Delta|V_{G,grid}|(t) = \int_{t_0}^t e^{\bA_v(t - \tau)}\bB_v\Delta Q_G \ \rmd \tau \quad \text{ (LIVE model)}.  \label{LIVE soln}
\end{align}
Moreover, the matrices $\bA_f$ and $\bA_v$ are diagonalizable, so the solutions \eqref{LIFE soln} and \eqref{LIVE soln} can be computed exceptionally quickly. 
\begin{comment}
A superior solving method for large, linear time-invariant (LTI) state equations involves reformulating (\ref{state_space}) as a discrete-time system using a matrix exponential as shown in (\ref{discretize}) \cite{dorfModernControlSystems2011}. The closed form solution is then (\ref{discretestatespace}). Assuming the generator dispatch is known, matrices $\mathbf{A_f}$ and $\mathbf{B_f}$ can be converted into their discrete-time formulations offline. Then, the disturbance magnitude and location, which is encoded in the vector $u$, serves as the input to (\ref{discretestatespace}). 
\\
%

\begin{equation}\label{discretize}
	\begin{split}
		A_d = e^{AT_s} \text{ },\quad B_d = (A_d- I)BA^{-1}
	\end{split}
\end{equation}
\begin{equation}\label{discretestatespace}
	x[k+1] = A_d x[k] + B_d u[k]
\end{equation}
\end{comment}

\begin{figure}[htbp]
	\centering \includegraphics[width=.75\columnwidth,trim={0 0 0 0},clip]{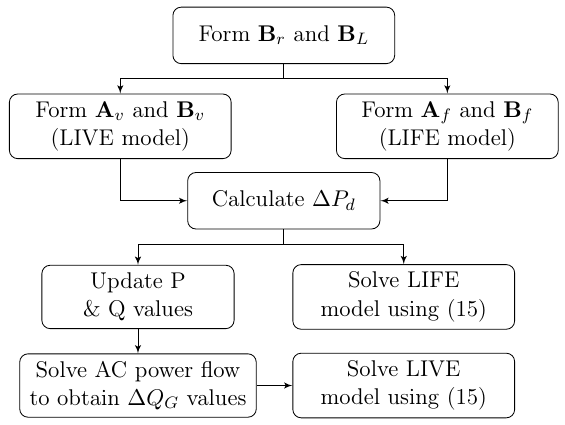}
	\caption{Workflow for the LIFE and LIVE models. The disturbance magnitude is embedded into $\Delta P_d$ as indicated in (\ref{delta_p_gen}).}
	\label{fig:flowchart}
\end{figure}

%
%\subsection{Implementation and Workflow}
%
The suggested workflow for the implementation of the LIFE and LIVE models is shown in Fig. \ref{fig:flowchart}. The susceptance matrix, $\mathbf{B}$ can be obtained from the MATPOWER, PowerWorld, TSAT, PSSE, or PSLF file of the relevant system for offline applications (planning case) and directly exported from the state estimator for online applications (operations case). The generator dispatches are determined by power flow calculations if conducting multi-scenario offline dynamics studies and directly by the state estimator if conducting online dynamic assessment. Next, for a given generator dispatch, $\bA_f$, $\bB_f$, $\bA_v$, and $\bB_v$ are calculated. The exact analytical solution is computed using either \eqref{LIFE soln} or \eqref{LIVE soln}. These computations are exceptionally fast, making them suitable to run on most commonly available computers for operational applications, real-time contingency screening, or for determining likely undesirable operating scenarios that may be investigated further with EMT simulation.

\section{Computer Simulation and Validation}\label{validation}

\begin{figure}[h]
	\centering
	\includegraphics[width=1\linewidth]{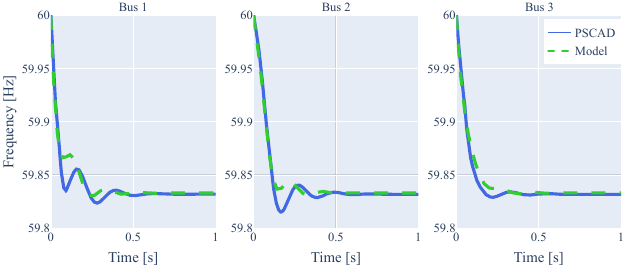}
	\caption{LIFE vs. PSCAD simulations in the 9-bus system.}
	\label{fig:9busvalidation}
\end{figure}

We used the WSCC 9-bus system and IEEE 39-bus system benchmarks to test and verify the accuracy of the frequency response model given in  (\ref{state_space}) against  industry-grade standard EMT simulations. The WSCC 9-bus model was constructed in PSCAD and utilizes the inverter models from \cite{kenyonOpenSourcePSCADGridFollowing2021}, which are quite complex and validated against  real-world data from the Maui power systems. In this system, all GFMs have a rated capacity of 200 MVA, an active power droop constant of $5\%$, and a power measurement time constant of 0.0628 seconds. A 31.5 MW + 9 MVAR load step at Bus 6 was simulated. A marginal loss constant of 1.035 is assumed. Visual inspection of the results indicate that the data from PSCAD verifies the accuracy of the LIFE model, as shown in Fig. \ref{fig:9busvalidation}. The runtimes to obtain five seconds of simulation data are given in Table \ref{tab:speeds}. To further demonstrate the computational advantage of our method, we also ran another version of the WSCC 9-bus system in PowerWorld and used the REGFM\_A1 model \cite{DuPositiveSequenceModeling}. The proposed model has the benefit of significantly accelerated solve time. While this is a small test case, computational accelerations of $\approx$ 9,219x over EMT simulations and $\approx$ 3,384x over RMS simulations create incredible opportunities for online operational applications.

\begin{table}[h]
	\centering   \caption{Simulation runtime comparison of the WSCC 9 bus network.}
	\begin{tabular}{c|ccc}
		& PSCAD & PowerWorld & LIFE Model\\\hline
		Runtime (sec) & 16.5940 & 6.0904 & 0.0018\\
		Acceleration factor & 9,219 & 3,384 & n/a\\
	\end{tabular}
	
	\label{tab:speeds}
\end{table}

\begin{figure}[htpb]
	\hspace{4pt}
	\begin{subfigure}[b]{0.235\textwidth}
		\includegraphics[width=\textwidth]{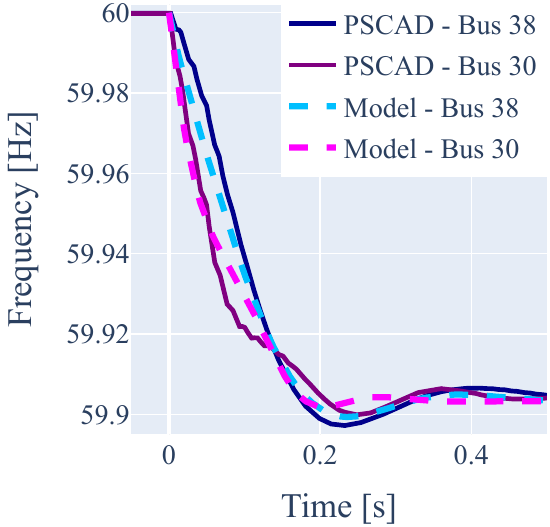}
		\caption{Buses 30 and 38.}
		\label{fig:39}
	\end{subfigure}
	\hfill
	\begin{subfigure}[b]{0.23\textwidth}
		\includegraphics[width=\textwidth]{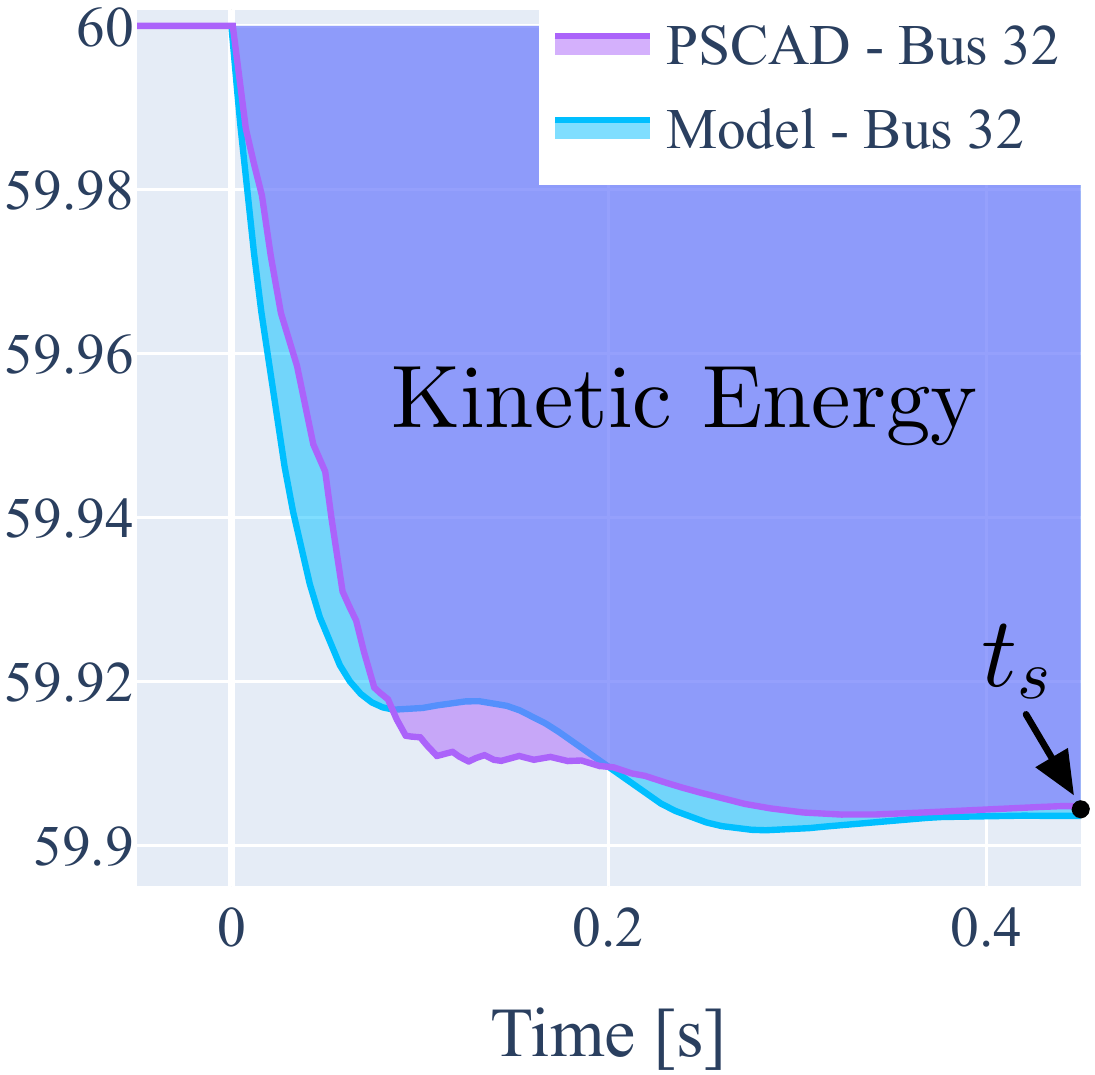}
		\caption{Bus 32 kinetic energy.}
		\label{fig:ke}
	\end{subfigure}
	\caption{LIFE vs. PSCAD results in the 39-bus system. The HS metric is a proxy for kinetic energy.}
	\label{fig:39busvalidation}
\end{figure}

The LIFE model was also validated against the 39 Bus system following a 307.5 MW + 141 MVAR load step at Bus 15. We observe good agreement between the LIFE model and the simulated trajectories, as shown in Fig. \ref{fig:39busvalidation}. In addition to nadir and RoCoF, we use the hertz-sec (HS) metric, which is a proxy for the kinetic energy resulting from a disturbance, to evaluate the accuracy of the LIFE model. The HS metric integrates the absolute value of frequency deviation over the transient period, as shown by the shaded area in Fig. \ref{fig:ke}. It is calculated using $HS = \int_{t_0}^{t_s}|f_0 - f(t)|dt$, where $t_0$ and $t_s$ are the time transient behavior begins and the time settling frequency is reached, respectively \cite{sajadiSynchronizationElectricPower2022}. The pre-disturbance frequency is $f_0$ and the transient frequency is a function of time, $f(t)$. In Table \ref{tab:39freq}, the error of nadir, RoCoF, and HS calculated using the LIFE model are given. While error of these key metrics was small, the required accuracy will be application-specific and as with any reduced-order model, there is a chance of estimations that are inconsistent with those of high-order models.

\begin{table}[htpb]
	\centering
	\caption{$|\text{Error}|$ of key metrics in the 39-bus system: LIFE vs. PSCAD.}
	\begin{tabular}{c|c|c|c}
		&\textbf{Nadir [Hz]}&\textbf{RoCoF [Hz/s]}&\textbf{HS [Hz$\cdot$s]}\\
		\hline
		Bus 30 &0.0079 &0.0518  &0.0006 \\
		Bus 31 &0.0042 &0.2420  &0.0002 \\
		Bus 32 &0.0041 &0.3504  &0.0003 \\
		Bus 33 &0.0046 &0.2915  &0.0004 \\
		Bus 34 &0.0088 &0.2693  &0.0001 \\
		Bus 35 &0.0048 &0.3523  &0.0006 \\
		Bus 36 &0.0068 &0.3790  &0.0006 \\
		Bus 37 &0.0075 &0.1668  &0.0003 \\
		Bus 38 &0.0106 &0.2381  &0.0000 \\
		Bus 39 &0.0088 &0.0712  &0.0009 \\ \hline
		Average &0.0068 &0.2412   &0.0004  \\
	\end{tabular}
	\label{tab:39freq}
\end{table}

\begin{figure}[h]
	\centering
	\includegraphics[width=.85\linewidth]{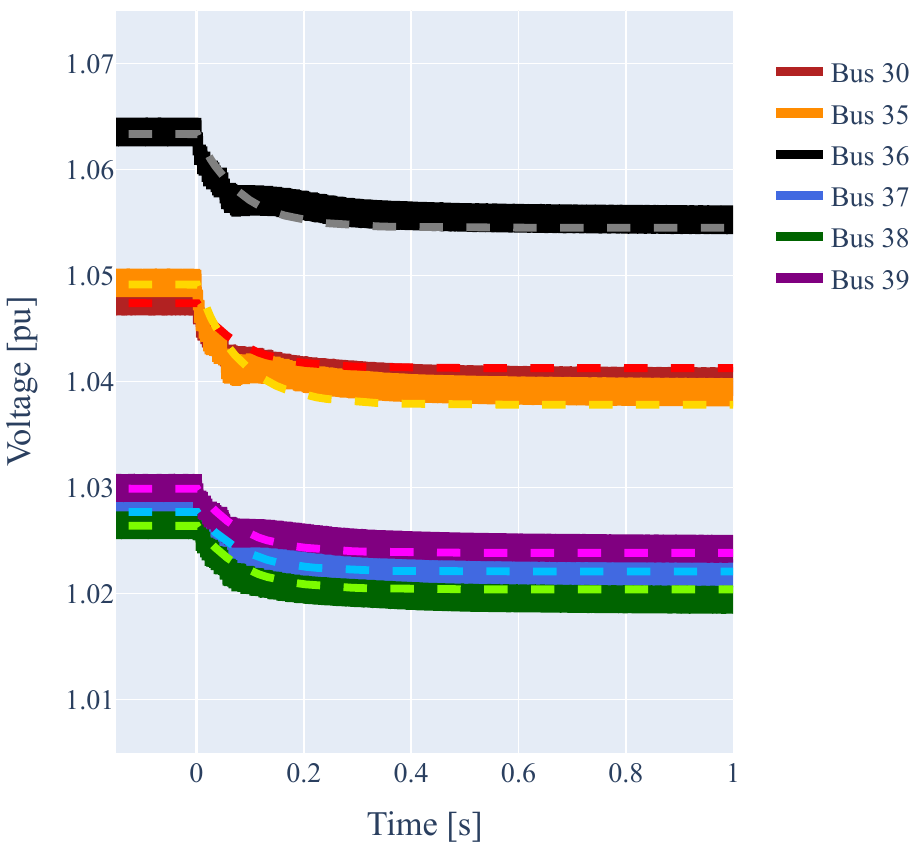}
	\caption{LIVE (dashed lines) vs. PSCAD simulations (solid lines) in the IEEE 39 bus system.}
	\label{fig:voltage_traces}
\end{figure}

Next, we validate the LIVE model against the same simulation scenarios (10\% complex load step) used to generate the traces in Figs. \ref{fig:9busvalidation} and \ref{fig:39busvalidation}. A $K_v$ value of 5 was used for both the 9 and 39 bus system. The results, shown in Table \ref{tab:9voltages}, Fig. \ref{fig:voltage_traces}, and Table \ref{tab:39voltages} point to the effectiveness of this approach. The PSCAD voltage traces in Fig. \ref{fig:voltage_traces} indicate negligible voltage oscillations, and therefore voltage transients are well approximated by a first order model without feedback. The lack of oscillation in the voltage magnitudes confirms that local changes in voltage magnitude and reactive power do not propagate well, and have a negligible effect on voltage magnitude at surrounding generator buses. While the inner-loop controls and LCL filter have an effect on the change in voltage magnitude at the generator bus, reasonable estimates of $\Delta Q_G$ and $K_v$ provide good approximations of voltage response without the need for explicit modeling of inner-loop controls. The validity of this approach is supported by the exceedingly small average error of settling voltage, which was just 0.0019 pu in the 9 bus system and 0.0021 pu in the 39 bus system. In the 39 bus system, the largest change in voltage magnitude was -0.011 pu, which occurred at Bus 35. However, like the proposed frequency response model, the LIVE model is a reduced-order model which carries some risk of inaccurate voltage estimations.
\begin{table}[htpb]
	\centering
	\caption{Settling voltage error [pu] in the 9-bus system: LIVE model vs. PSCAD simulations. }
	\begin{tabular}{c|ccc}
		& Bus 1 &Bus 2  &Bus 3  \\ \hline
		Error [pu]&  0.0001 &0.0043 & 0.0015 \\
	\end{tabular}
	\label{tab:9voltages}
\end{table}
\begin{table}[htpb]
	\centering
	\caption{Settling voltage error [pu] in the 39-bus system: LIVE model vs. PSCAD simulations. }
	\begin{tabular}{c|cccccc}
		& Bus 30 &Bus 31  &Bus 32  &Bus 33  &Bus 34  &Bus 35 \\ \hline
		Error [pu] & 0.0021 & 0.0012 & 0.0014 & 0.0009 & 0.0029 & 0.0003 \\
		&Bus 36  &Bus 37  &Bus 38  &Bus 39\\
		\hline
		Error [pu] & 0.0030 & 0.0032 & 0.0045 & 0.0014 &  &  \\
	\end{tabular}
	\label{tab:39voltages}
\end{table}
\section{Scalability and Application: 25,000-bus System}\label{application}
Having demonstrated the efficacy of the proposed models in the previous section, we now focus on their scalability and provide a discussion on potential applications.
\subsection{Scalability}
\begin{figure}[h]
	\centering \includegraphics[width=.8\columnwidth,trim={0 0 0 0},clip]{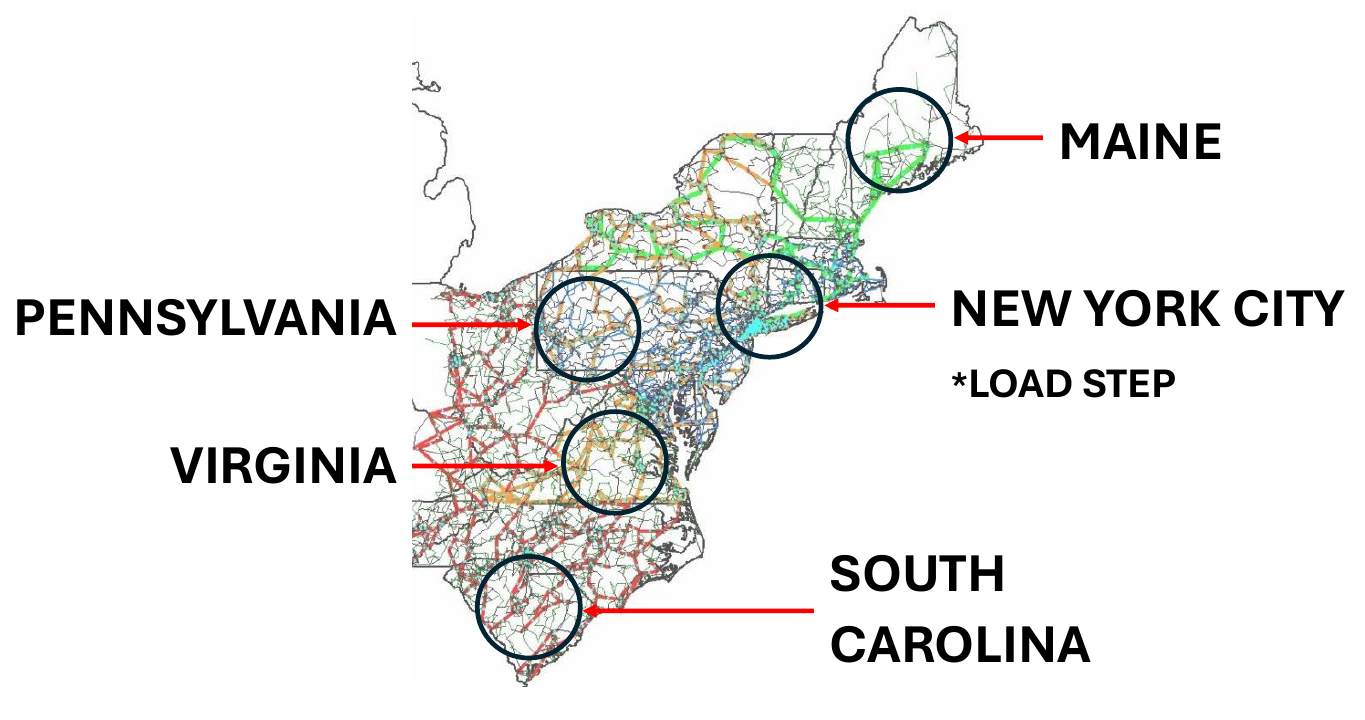}
	\caption{25k bus synthetic grid of the northeastern United States \cite{BirchfieldGridStructural}. A load step was applied in New York City, and frequency was observed in five regions across the network.}
	\label{fig:25kpic}
\end{figure}

\begin{figure}[h]
	\centering \includegraphics[width=1\columnwidth,trim={0 0 0 0},clip]{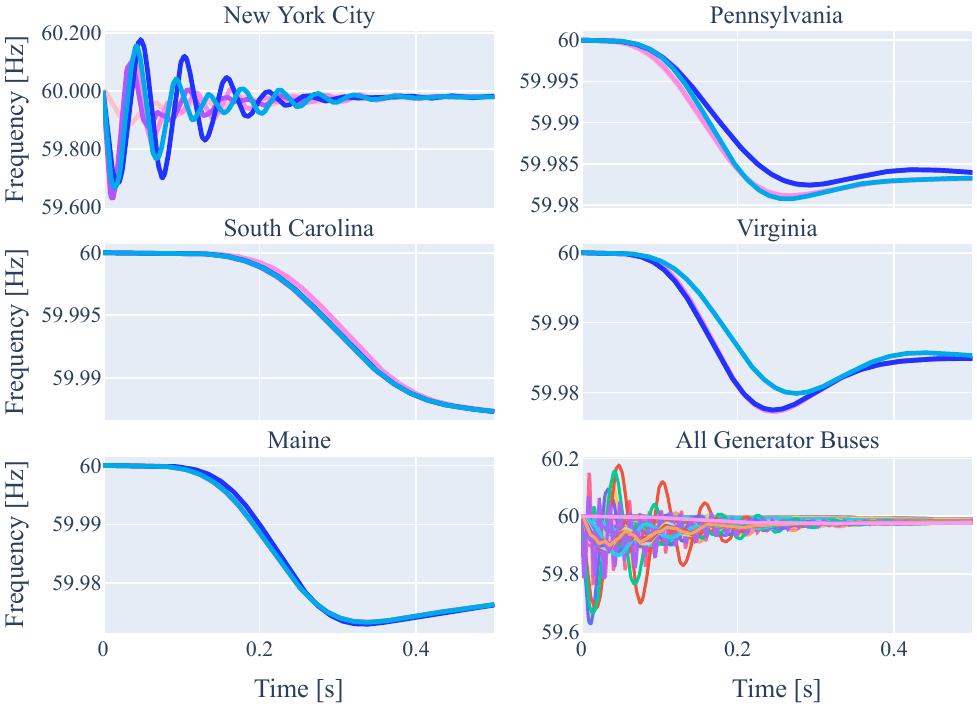}
	\caption{Predicted frequency response in five regions across the ACTIVSg25k network. Frequency traces at all 849 GFM buses are shown in the bottom right corner. All frequencies reached a stable value of $59.98 \pm  0.005 \text{ Hz}$ after $0.589$ seconds.}
	\label{fig:25kresults}
\end{figure}

 The frequency response and voltage response models were applied to the open-source ACTIVSg25k synthetic grid of the northeastern United States \cite{BirchfieldGridStructural}, which is shown in Fig. \ref{fig:25kpic}. This is an extremely large system, comprising most of three independent system operators' territories, with additional vertically integrated utility territory included. For every bus with multiple generators, we aggregated generators into one generator with an equivalent capacity. We then converted all the generators with capacities exceeding $100 \text{ MVAR}$ to be droop-controlled GFMs. This resulted in $849$ GFM buses. The remaining $24,151$ buses are modeled as load buses in (\ref{networkeqn}). All generators below this capacity were considered GFLs and modeled as negative load. A very large system formed entirely by GFMs is an exaggerated version of potential future power systems. We utilize the frequency and voltage models on such a system to demonstrate their efficiency on systems where EMT simulation would be impossible and computational efficiency is paramount. A $1.56 \text{ GW} + 780\text{ MVAR}$ load step was applied to Bus $69665$ in New York City. The active power disturbance magnitude is equal to the capacity of the largest online generator, and the reactive power disturbance magnitude is equal to half of the active power disturbance magnitude. The active-droop constant and power measurement time constant of each inverter were $5\%$ and $0.0628$ seconds, respectively. The solution, given in (\ref{LIFE soln}), was solved in only $1.877$ seconds to obtain $1$ second worth of simulated frequency response, which is sufficient given the faster response of GFMs compared to SGs. To illustrate the heterogeneity of frequency response, the predicted frequency response in five different areas across the system are shown in Fig. \ref{fig:25kresults}. The frequency trajectories for all GFM buses are also shown. The results indicate that while all frequencies in the system synchronize within a second, GFM buses closest to the disturbance experience dramatic frequency oscillations and more distant buses display much smoother trajectories. This behavior is expected, seeing as GFM buses closest to a disturbance will be subjected to greater changes in active power injection, according to (\ref{delta_p_gen}). Also noteworthy is the time delay between the onset of transients between areas close to the disturbance versus areas farther from the disturbance. This is perhaps most obvious when comparing the frequency trajectories in New York City to those in South Carolina, where frequency deviation is delayed by about $.18$ seconds. This heterogeneity in frequency response trajectories across the system would not have been seen if one were using a SFR model to represent the system with a single trajectory.

\begin{figure}[h]
	\centering \includegraphics[width=.9\columnwidth,trim={0 0 0 0},clip]{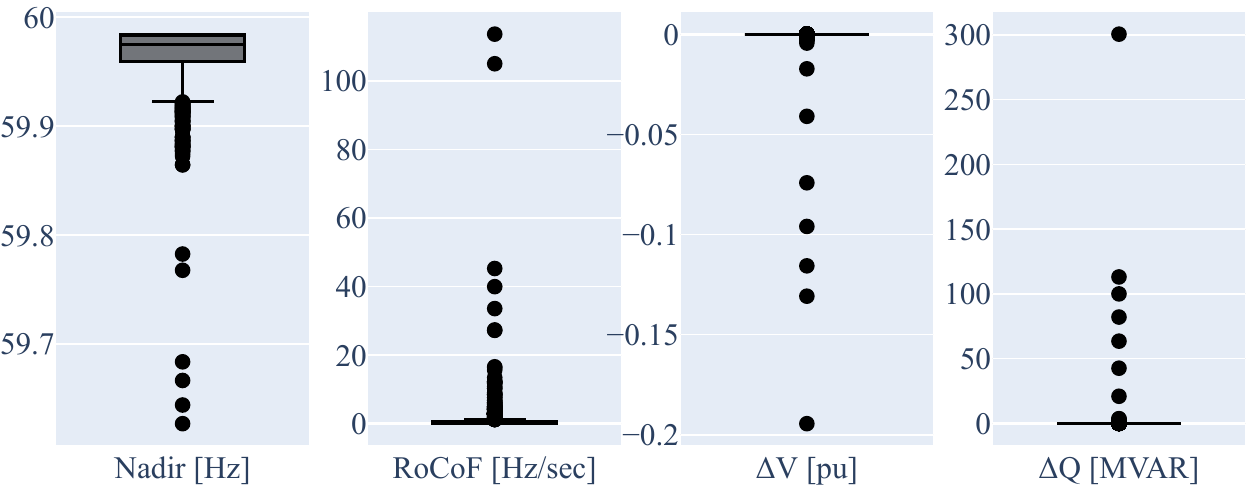}
	\caption{Box plots of nadir, absolute value of RoCoF, change in voltage magnitude, and change in reactive power injection at each GFM bus in the ACTIVSg25k test case.}
	\label{fig:boxes}
\end{figure}

Fig. \ref{fig:boxes} depicts the box plots of the LIFE nadir and RoCoF values in the ACTIVSg25k network post-disturbance. The LIFE model captures the spatial and temporal differences in frequency response as these variations in frequency could mean the difference between protection mechanisms tripping or not. Therefore, the model's ability to capture frequency response extrema is exceedingly important. In this scenario, the average frequency nadir was $59.96 \text{ Hz}$ and the average RoCoF was $1.30 \text{ Hz/sec}$. Of the 849 GFM buses, $143$ buses experienced RoCoFs exceeding $1 \text{ Hz/sec}$ and the maximum RoCoF was a staggering $113.6 \text{ Hz/sec}$. As expected, the largest RoCoF values were seen at buses closest to the disturbance and at buses with small GFM capacities. Even so, the nadir frequency at these buses is about  $59.77 \text{ Hz}$, which does not violate the underfrequency load shedding limit \cite{Manual33System2024}. 

\begin{figure}[h]
	\centering \includegraphics[width=01\columnwidth,trim={0 0 0 0},clip]{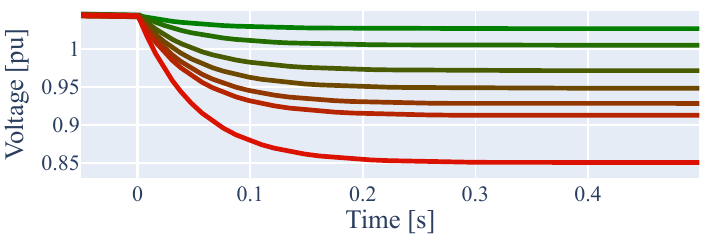}
	\caption{Voltage trajectories at GFM buses where the change in voltage magnitude exceeded $0.01 \text{ pu}$.}
	\label{fig:25kresultsV}
\end{figure}

The voltage response model solved in just $0.127$ seconds due to the approximation of Q-V dynamics as an open-loop process. $K_v$ was set equal to $5$. As seen in Fig. \ref{fig:boxes}, most GFM buses experienced negligible changes in reactive power injection and voltage magnitude, with a few crucial exceptions. The average change in voltage magnitude was  $0.0009 \text{ pu}$ and the maximum was $0.19 \text{ pu}$. The predicted voltage trajectories at GFM buses where $\Delta V > 0.01 \text{ pu}$ are shown in Fig. \ref{fig:25kresultsV}. The greatest changes in reactive power injection and voltage variations are seen at the generators electrically closest to the load step location, which makes sense given the limited range of reactive power disturbances. There were four GFM buses at which reactive power deficiency caused the voltage magnitude to settle below $0.95 \text{ pu}$.

\subsection{Applications}
The LIFE and LIVE models demonstrate promising utility for several applications to enhance the efficiency of power system studies; here we expand on three of them. 

At present, methods for screening risks to dynamic stability are largely limited to techniques based on the short-circuit ratio (SCR) concept \cite{GroganEMT}. These SCR-based methods are limited in that they do not yield post-disturbance frequency and voltage trajectories, rather they are simply values meant to indicate where in a system stability issues may arise. These methods also assume identical IBRs and do not distinguish between control system designs \cite{GroganEMT}. Seeing the need for screening methods that reveal more about the transient behavior of a system than SCR-based methods offer, the LIFE and LIVE models may be used  prior to full-order EMT simulation in order to identify dynamically interesting operational scenarios and areas within a system most vulnerable to instability. In this application, the models would provide useful information regarding the predicted severity of a disturbance in terms of voltage and frequency stability for different dispatch and generation location scenarios.

Another potential application of the proposed models is for parametric analysis. Conducting this type of analysis using EMT tools is difficult because it requires many computationally expensive simulations. The significant speed advantage of the LIFE and LIVE models over EMT simulation makes their application in parametric sensitivity analysis attractive, as does the ability to derive participation matrices using the LIFE and LIVE state-space representations. By using the proposed models, analyzing the transient behavior of a system under a wide range of parametric scenarios and mathematically drawing connections between states and the dominant modes of a system may be done rapidly. 

Finally, the proposed models may be integrated into a variety of long-range planning tools, including capacity expansion and resource adequacy studies. Rather than only conducting a limited number of stability studies with EMT simulation tools based on the results of an optimization problem, the LIFE and LIVE models could be incorporated directly into an optimization framework, allowing systems planners to introduce both frequency and voltage stability as constraints. By integrating the proposed models into an optimization framework as described instead of relying solely on EMT simulation, power system operators can greatly reduce the number of required EMT simulations.

\section{Conclusion}\label{conclusion}
While frequency response models have a history spanning decades, many models disregard the heterogeneity of the frequency response and assume that the network is dominated by synchronous generators. All disregard reactive power-voltage dynamics. In this paper we present models of frequency and voltage response for 100\% IBR networks. The models were validated on the WSCC 9 bus and IEEE 39 bus systems. The models were then utilized to predict frequency and voltage response in a 25,000 bus synthetic system. Our results indicate that the models provide reasonable estimations of frequency and voltage response at every GFM bus in the system while, crucially, maintaining computational tractability, setting the basis for their use in planning and operational applications.

\section{Acknowledgments}
We would like to thank Dr. Mostafa Sedighizadeh, Mr. Harvey Scribner, and Mr. Nick Parker at Southwest Power Pool (SPP), Dr. Mohammadi Mohammadi of the Australian Energy Market Operator (AEMO), and Dr. Bruno Leonardi of the New York Independent System Operator (NYISO) for their insightful feedback and constructive comments.

This material is based upon work supported by the National Science Foundation Graduate Research Fellowship under Grant No. DGE $2040434$ and in part by the National Science Foundation’s Advancing Sustainability through Power Infrastructure for Roadway Electrification (ASPIRE) Engineering Research Center and National Science Foundation, United States Award $1941524$.

This work was authored in part by the National Renewable Energy Laboratory, operated by Alliance for Sustainable Energy, LLC, for the U.S. Department of Energy (DOE) under Contract No. DE-AC$36$-$08$G$O28308$. The views expressed in the article do not necessarily represent the views of the DOE or the U.S. Government. The U.S. Government retains and the publisher, by accepting the article for publication, acknowledges that the U.S. Government retains a nonexclusive, paid-up, irrevocable, worldwide license to publish or reproduce the published form of this work, or allow others to do so, for U.S. Government purposes.

\vspace{-2cm}
\begin{IEEEbiography}
[{\includegraphics[width=1in,height=1.25in]{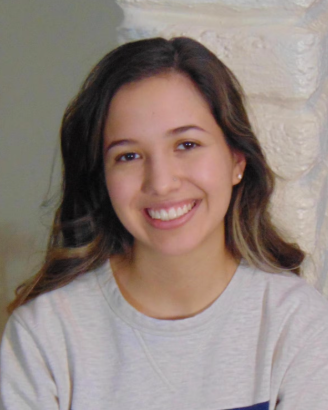}}]{Marena~Trujillo} (Student Member)
is a PhD student in Electrical, Computer \& Energy Engineering at the University of Colorado Boulder, Boulder, CO, USA. 
She is a NSF GRFP Fellow, and her main research interests include the dynamics and stability of low-inertia inverter-based power systems, and the use of numerical methods and model reduction techniques to enable efficient, high-fidelity simulation of large
systems.
In 2021, she received her B.S.E. in electrical engineering  from Loyola Marymount University, Los Angeles, CA, USA.
\end{IEEEbiography}

\vspace{1.5cm}

\begin{IEEEbiography}[{\includegraphics[width=1in,height=1.25in]{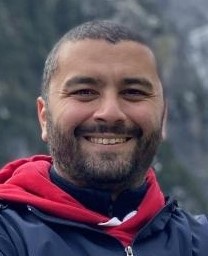}}]{Amir~Sajadi} 
(S'12, M'16, SM'19) 
is an Associate Research Professor with the Renewable and Sustainable Energy Institute, University of Colorado Boulder, Boulder, CO, USA, and a Research Affiliate with the National Renewable Energy Laboratory (NREL), Golden, CO, USA. 
He received the Ph.D. degree in systems and control engineering from Case Western Reserve University, Cleveland, OH, USA. His main research interests include modeling, planning, dynamics, control, and management of large-scale power systems to enhance the stability, security, and resiliency of energy delivery. 
\end{IEEEbiography}
\vspace{-10.5cm}

\begin{IEEEbiography}[{\includegraphics[width=1in,height=1.25in]{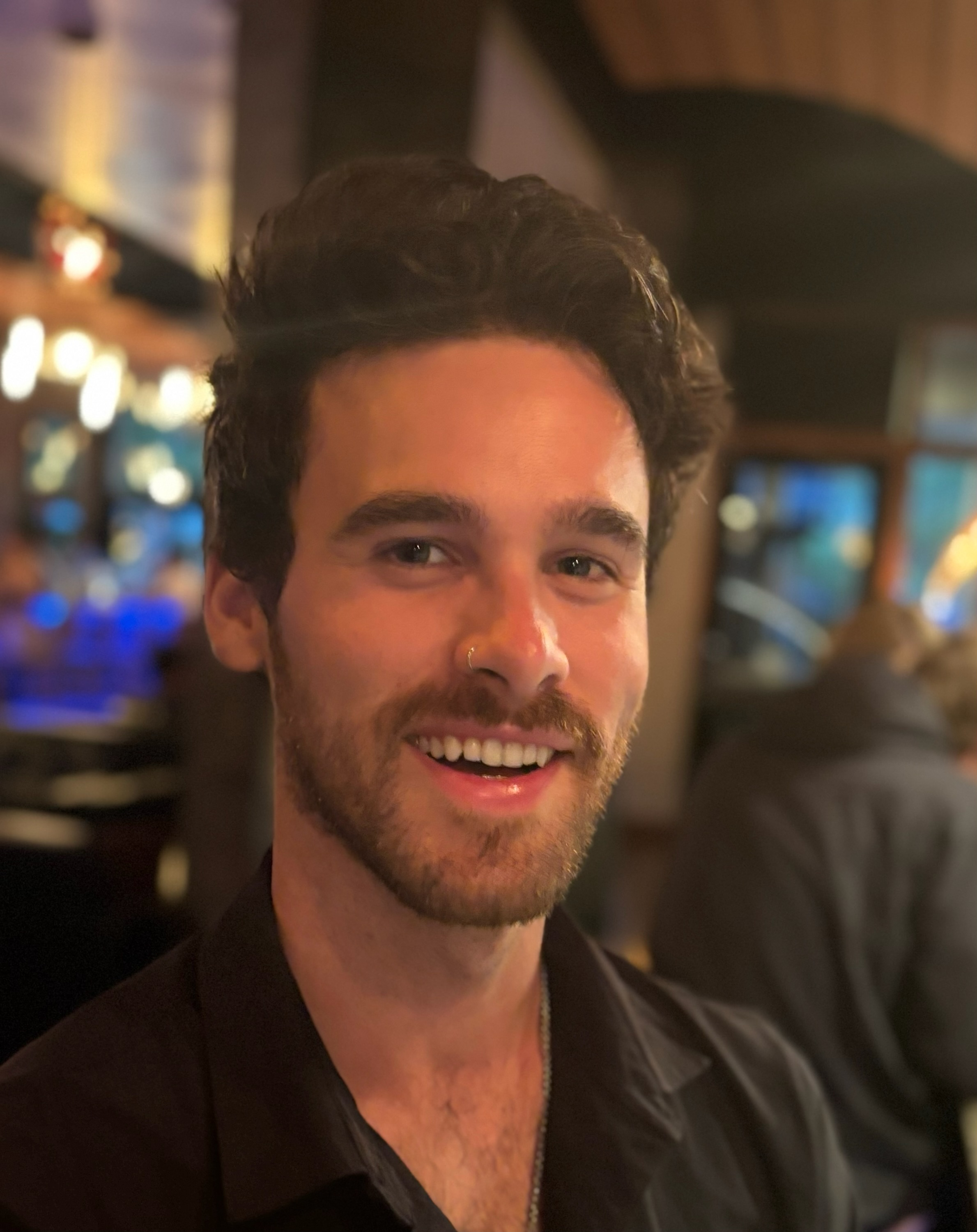}}]
{Jonathan~Shaw} is a PhD student in the Applied Mathematics department at the University of Colorado Boulder, Boulder, CO, USA. His research interests include applied functional analysis, numerical analysis, and asymptotic methods in the field of low-inertia power systems. He received his B.S. in applied mathematics, with a concentration in aerospace engineering, from the aforementioned university in 2023. 
\end{IEEEbiography}
\vspace{-10.5cm}

\begin{IEEEbiography}
[{\includegraphics[width=1in,height=1.25in,clip,keepaspectratio]{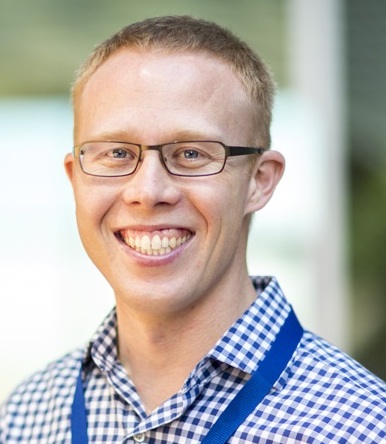}}]
{Bri-Mathias Hodge} 
(M'10, SM'17) is a Professor with the Department of Electrical, Computer and Energy Engineering and Associate Director and a Fellow of the Renewable and Sustainable Energy Institute, University of Colorado Boulder. He previously was Chief Scientist and Distinguished Member of Research Staff with the Power Systems Engineering Center, National Renewable Energy Laboratory (NREL). He received the B.S. from Carnegie Mellon University, M.S. from \AA bo Akademi University, and Ph.D. from Purdue University.
\end{IEEEbiography}

\end{document}